\documentclass[prx,10pt,amsmath,amssymb,letterpaper,nobalancelastpage,aps,showpacs,preprintnumbers,twocolumn]{revtex4-1}
\usepackage{graphicx}
\usepackage{dcolumn}
\usepackage{bm}
\usepackage{slashed}
\usepackage{amssymb}
\usepackage{amsmath}
\usepackage{amsthm}
\usepackage{color}

\begin{document}

\title{Phase diagram of two  interacting helical states}

\author{Raul A. Santos$^{1,2}$, D.B. Gutman$^{1}$ and Sam T. Carr$^{3}$}
\affiliation{$^{1}$Department of Physics, Bar-Ilan University, Ramat Gan, 52900, Israel}
\affiliation{$^{2}$Department of Condensed Matter Physics, Weizmann Institute of Science, Rehovot 76100, Israel}
\affiliation{$^{3}$School of Physical Sciences, University of Kent, Canterbury CT2 7NH, United Kingdom}


\begin{abstract}
We consider two coupled time reversal invariant helical edge modes of the same helicity, such as would occur on two stacked quantum spin Hall insulators.  In the presence of interaction, the low energy physics is described by two collective modes, one corresponding to the total current flowing around the edge and the other one describing relative fluctuations between the two edges.  We find that quite generically, the relative mode becomes gapped at low temperatures, but only when tunneling between the two helical modes is non-zero.  There are two distinct possibilities for the gapped state depending on the relative size of different interactions.  If the intra-edge interaction is stronger than the inter-edge interaction, the state is characterised as a spin-nematic phase.   However in the opposite limit,
when the interaction between the helical edge modes is strong compared to the interaction within each mode,
a spin-density wave forms, with emergent topological properties.  Firstly, the gap protects the conducting phase against localization by weak nonmagnetic impurities; and secondly the protected
phase hosts localized zero modes on ends of the edge that may be created by sufficiently strong non-magnetic impurities.
\end{abstract}
\date{\today}
\preprint{}

\maketitle

\section{Introduction}
 Symmetry protected topological states of matter are characterized by the invariance of their Hamiltonian under local
 symmetries. These states are referred as topological insulators/superconductors. They possess gapless surface modes that are 
 protected by the gap in the bulk of the material, as long as the
 symmetries are not broken. For non-interacting particles the topological classification is determined by 
 time reversal (TR) and particle-hole (PH) symmetry \cite{Kitaev2009,Shinsei2010}. Under this classification, 
 a two dimensional insulator invariant under TR symmetry can be either trivial or topological. 
 While the bulk  conductivity vanishes at zero temperature in both cases,  a non-trivial topological insulator (TI)  
 hosts gapless helical edge modes\cite{Kane2005,Bernevig2006}. A single  helical edge mode consist of a Kramers-pair, 
 connected by TR symmetry. The disorder that does not break the TR symmetry can not scatter between the Kramers partners. 
Therefore the system is  protected against localization  as long as the gap in the bulk exceeds the disorder potential 
and TR symmetry is preserved.  The nontrivial TR topological insulators, also known as quantum spin Hall insulators (QSHI), have been observed 
 experimentally in certain two dimensional \cite{Konig2007,Konig2008,Hsieh2009} materials with strong spin-orbit.  An analogous state occurs in three dimensions\cite{Hsieh2008,Xia2009,Hsieh2009b}, where the two-dimensional surface is conducting and cannot be localised.

Such a state is known as a $\mathbb{Z}_2$ topological insulator, meaning that the number of protected topological modes is either zero or one.  This means that if one considers a systems with two helical edge modes, backscattering between non Kramers pairs 
is allowed, leading to  Anderson' localization of the  edge modes. In this case the system is a topologically trivial insulator.  Whether it is possible to find an individual material exhibiting two (non-protected) helical modes or not is, as far as we know, an open question.  However such a setup can certainly be engineered by considering a stack of two QSHIs, sufficiently close that the conducting edge modes may both hybridise and interact with each other via the Coulomb interaction.
 
 The presence of electron-electron interactions can dramatically change the 
 properties, even of single QSHIs \cite{Beri2012,Levin2012,Sela2011,Oreg2014}. In particular, as it has been shown in Ref. \cite{Schmidt2012,Kainaris2014}
the topological protection of a single  helical mode in the presence of impurities 
is removed  by sufficiently strong repulsive interactions.
The process  involves  coherent scattering of two interacting electrons off a static impurity, a process allowed by TR symmetry. As a result  the helical state is localized.  
On the other hand, as shown in Ref. \cite{Santos2015}  moderate repulsive  interaction stabilizes the conducting phase, for TI with a number of edge modes.  Clearly  these two mechanisms act in opposite direction.  In this work we  complete the analysis of \cite{Santos2015} and take into account two particle scattering.

We focus on  two  helical edge modes, coupled by tunneling and electron interaction \cite{Tanaka2009,Santos2015}. In the non-interacting limit this system is 
topologically equivalent to a trivial insulator. We show that in the presence of interaction  the system may or may not be topologically
trivial depending on the strength of interaction and tunneling amplitude. If the inter-mode interaction is smaller than the  interaction between  Kramers pairs, the system remains topologically trivial, with vanishing conductance at zero temperature. 
In this case, the system is in a spin-nematic phase \cite{Nersesyan1991,GogolinBook2004,FradkinBook2013}. 
In the opposite limit, where the inter-mode interaction is stronger than the interaction within each mode, 
the system remains conducting. This protection against localization is a direct consequence of the spin gap.
By adding strong non-magnetic impurities  the  edge mode 
splits into unconnected parts, each   hosting a pair of  localized zero modes on its ends. 

This paper is organized as follows. In the first section we formulate the model. 
In the second section we apply  bosonization technique and  analyze the low 
temperature fixed point using  the renormalization group (RG). 
In the third section we study  the stability of the conducting phase  against a  single impurity and random
disorder.  We summarize and discuss our results in the conclusion.

 \begin{figure}[t]
	\centering
		\includegraphics[width=\linewidth]{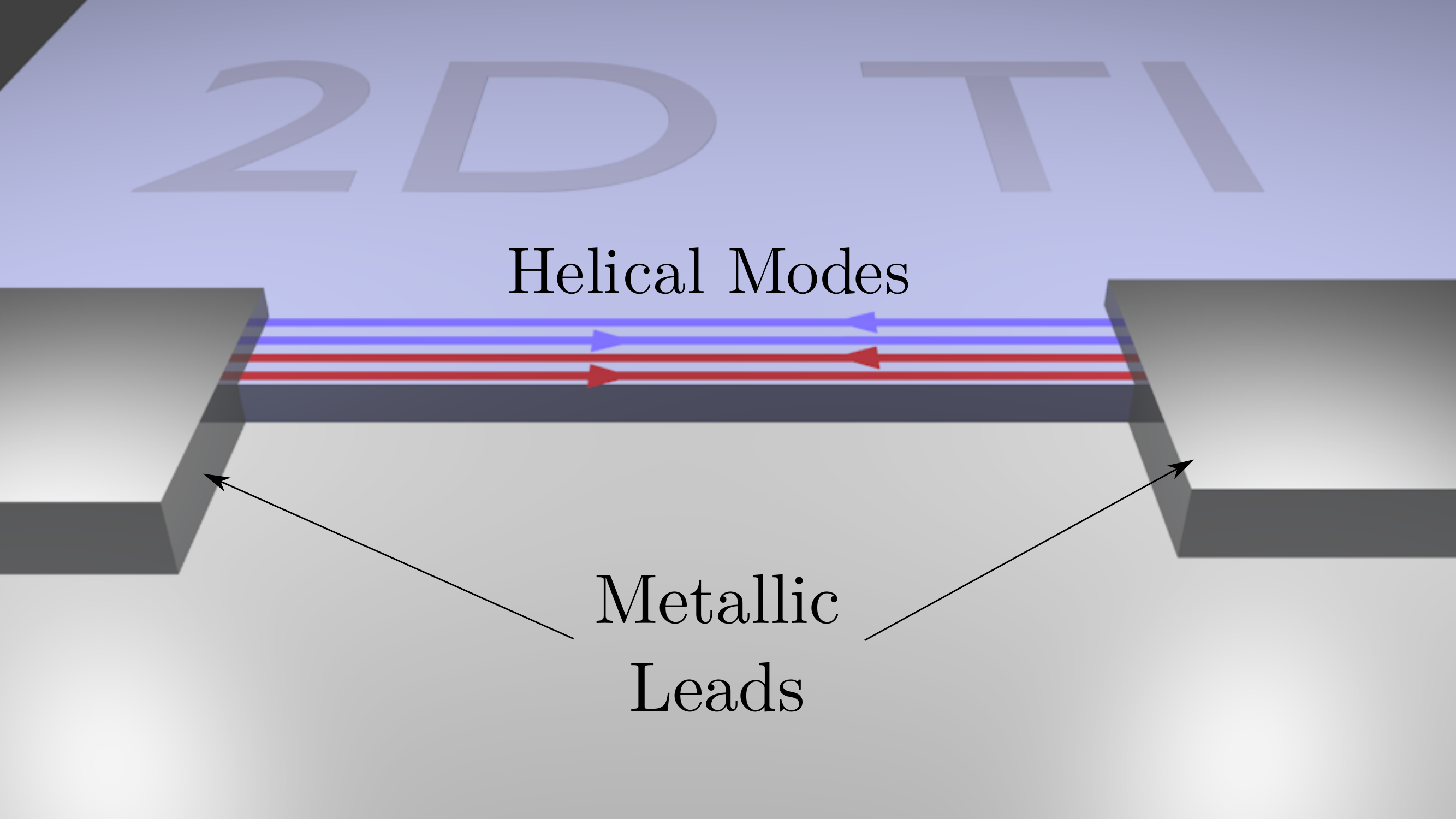}
	\caption{(color online) Experimental setup to measure the conductance (See text).
	}\label{fig:Exp_leads}
\end{figure}

\section{Two coupled helical modes}

We consider  two interacting helical modes. Each one  is formed  at the edge of 
a two dimensional TR invariant topological insulator that are placed one next to another, see Fig.\ref{fig:Exp_leads}
The Hamiltonian of the clean system (disorder or impurities will be added in Section \ref{sec:disorder}) consists of four different parts
\begin{equation}\label{H_0_nonint}
 H=H_{\rm kin}+H_{\rm tun}+H_{\rm SO}+H_{\rm int}.
\end{equation}
Here $H_{\rm kin}$ is the kinetic energy
\begin{equation}
 H_{\rm kin}=\sum_{k,\sigma,a}\epsilon_{\sigma,a}(k)c^\dagger_{\sigma,a}(k)c_{\sigma,a}(k),
\end{equation}
where $c^\dagger_{\sigma,a}(k)$   creates a fermion in a helical mode ($a=1,2$) with a given  
spin ($\sigma=\uparrow,\downarrow$)
and momentum $k$; $\epsilon_{\sigma,a}(k)$ is 
the dispersion relation of the noninteracting mode.  The helicity comes from the relationship between spin $\sigma$ and the dispersion -- roughly speaking spin up will correspond to a right moving mode while spin down will be a left moving mode; this will be fully discussed below.

The tunneling between the two modes is described by
\begin{equation}
 H_{\rm tun}= -t_\perp\sum_{k,\sigma,a}c^\dagger_{\sigma,a}(k)c_{\sigma,\bar{a}}(k),
\end{equation}
where we introduced the notation $\bar{1}=2,\bar{2}=1$. 

In a helical model when spin is related to chirality, for there to be any backscattering at all, one must break $S_z$ symmetry.  While many previous works did this at a phenomenological level (as TRS does not imply unbroken $S_z$ symmetry), we expand on a model originally proposed for a single edge by Schmidt \textit{et al} \cite{Schmidt2012}.  In this model, the spin-orbit coupling $H_{\rm SO}$ is explicitly incorporated into the non-interacting part of the model, as this is the physical process that leads to broken spin-rotation symmetry.
This coupling, $H_{\rm SO}$ has the generic form
\begin{equation}
 H_{\rm SO}=\alpha_{SO}\sum_{k,\sigma,\sigma',a}k c^\dagger_{\sigma,a}(k)(\sigma^x)_{\sigma,\sigma'}c_{\sigma',a}(k),
\end{equation}
with $\sigma^x$ being  the corresponding  Pauli matrix.

Finally, the  interaction between  electrons  is modeled by 
\begin{eqnarray}\label{interaction_ham}
 H_{\rm int}=U_0\sum_{x,a}n_a(x)n_a(x)+2U\sum_{x}n_1(x)n_2(x).
\end{eqnarray}
Here $U_0$ and $U$ stand  for interaction constants within the same mode and between different modes consequently. Under generic conditions these two constants are different  ($U_0\neq U$).
The fermion densities are 
\begin{equation}
n_a(x)=c^\dagger_{\uparrow,a}(x)c_{\uparrow,a}(x)+c^\dagger_{\downarrow,a}(x)c_{\downarrow,a}(x),
\end{equation}
where, as usual $c_{\sigma,a}(x)=\sum_k e^{ikx} c_{\sigma,a}(k)$. 

Throughout this work, we will use units where $\hbar=1$ and $a_0$ is the short-distance cutoff for the field theory.  This may be thought of as an effective lattice spacing for the helical modes; however in a full theory of the entire two-dimensional setup of the QSHI, it is more closely related to the inverse of the bulk gap.  In either case, it is a non-universal constant in the field theory that sets the overall energy scale.

\subsection{Diagonalization of Non-interacting Hamiltonian}

\begin{figure}
	\centering
		\includegraphics[scale=0.4]{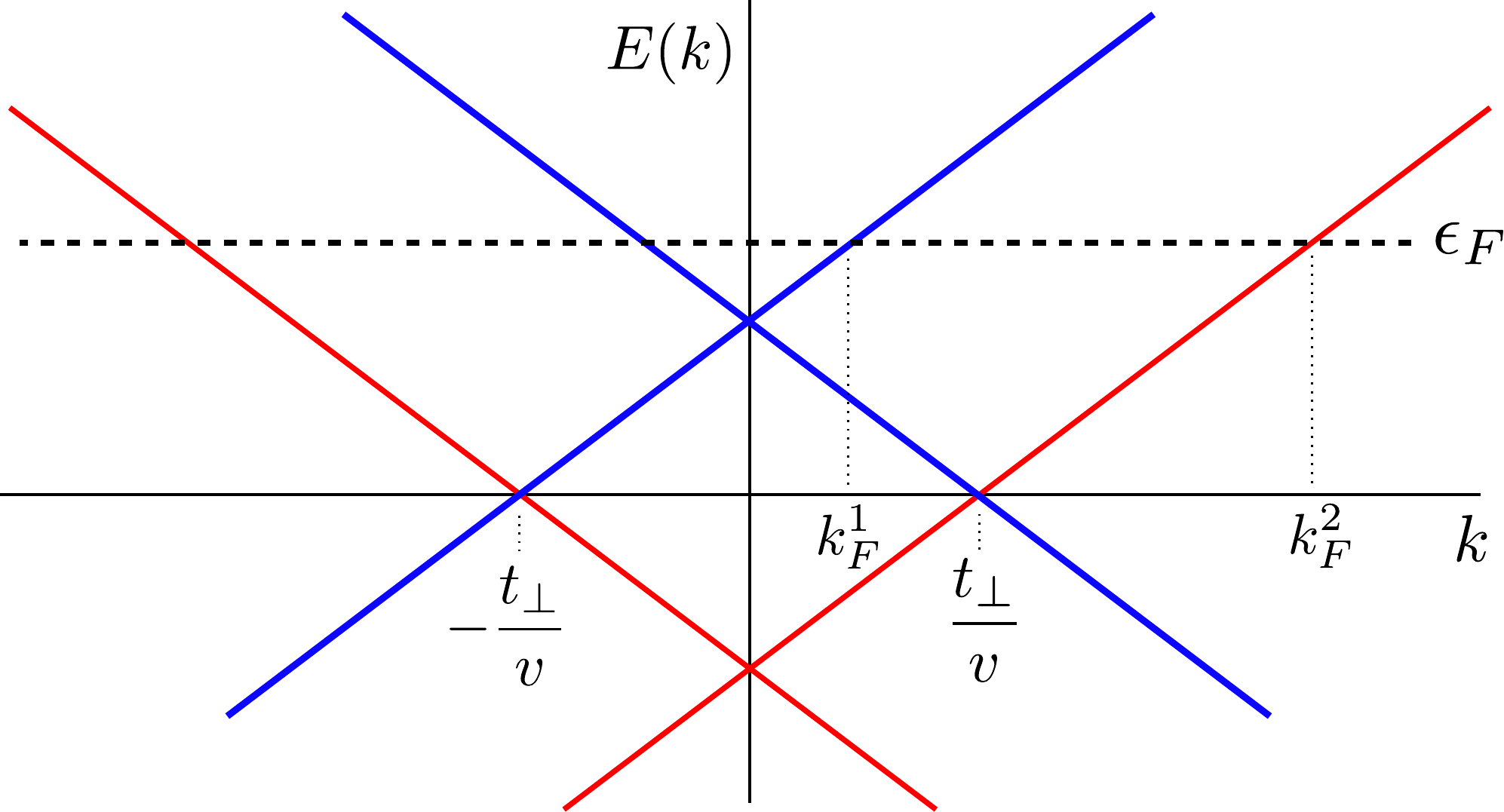}
	\caption{(color online).- Single particle energy spectrum. Kramers pairs are depicted by  the same color. }\label{fig:SPE}
\end{figure}

In a TR invariant system, the dispersion relation must satisfy  the constraint
\begin{equation}
 \epsilon_{\sigma,a}(k)=\epsilon_{\bar{\sigma},a}(-k).
\end{equation}
The simplest dispersion relations describing gapless modes are $\epsilon_{\uparrow,a}(k)=v_F k$, and
$\epsilon_{\downarrow,a}(k)=-v_F k$. Here we assume that the Fermi velocity of the non interacting helical modes
is the same. Introducing the vector of fermionic fields 
\begin{equation}
 \boldsymbol{c}^\dagger(k)=(c^\dagger_{\uparrow,1}(k),c^\dagger_{\downarrow,1}(k),c^\dagger_{\uparrow,2}(k),c^\dagger_{\downarrow,2}(k)),
\end{equation}
the noninteracting part of the Hamiltonian $H_0=H_{\rm kin}+H_{\rm tun}+H_{\rm SO}$ becomes
\begin{equation}
\label{h0}
 H_0=\sum_{k}\boldsymbol{c}^\dagger(k)h_0(k)\boldsymbol{c}(k),
\end{equation}
with $h_0(k)$ the Hermitian matrix
\begin{eqnarray}\label{matrix_non_int}
 h_0&=&\begin{bmatrix}
     v_F k & \alpha_{SO} k & -t_\perp & 0 \\
     \alpha_{SO} k & -v_F k  & 0 & -t_\perp  \\
     -t_\perp & 0  & v_F k & \alpha_{SO} k \\
      0 & - t_\perp & \alpha_{SO} k  & -v_F k
 \end{bmatrix},\\\label{matrix_non_int2}
 &=&\delta_{aa'}(v_F\sigma^z_{\sigma\sigma'}+\alpha_{SO}\sigma^x_{\sigma\sigma'})k-t_\perp\tau^x_{aa'}\delta_{\sigma\sigma'}.
\end{eqnarray}
Here $\sigma^{x,y,z},\tau^{x,y,z}$ are the corresponding Pauli matrices in spin and mode space respectively. Using (\ref{matrix_non_int2}) 
we find that $h_0(k)$ is diagonalized by an unitary transformation $B$, such that $h_0(k)=B^\dagger D(k) B$, with 
\begin{eqnarray}\label{matrix_rotation}
 B&=&\frac{1}{\sqrt{2}}\begin{bmatrix}
     \cos\beta & \sin\beta  & \cos\beta & \sin\beta \\
     -\sin\beta & \cos\beta  & -\sin\beta & \cos\beta \\
     \cos\beta & \sin\beta  & -\cos\beta & -\sin\beta \\
     -\sin\beta & \cos\beta  & \sin\beta & -\cos\beta \\
 \end{bmatrix}\\
 &=&\frac{(\tau^z_{aa'}+\tau^x_{aa'})}{\sqrt{2}} (e^{i\beta\sigma^y})_{\sigma\sigma'},
\end{eqnarray}
and  $\beta=\frac{1}{2}{\rm tan}^{-1}\left(\frac{\alpha_{SO}}{v_F}\right).$
We therefore pass to the new basis
\begin{equation}
\label{psi}
 \boldsymbol{\psi}(k)=B\boldsymbol{c}(k),\,\,\,
\boldsymbol{\psi}^\dagger=(\psi^\dagger_{+,1},\psi^\dagger_{-,1},\psi^\dagger_{+,2},\psi^\dagger_{-,2}), 
\end{equation}
where the  single particle Hamiltonian (\ref{h0})  is diagonal
\begin{equation}
 H_0=\sum_{k}\boldsymbol{\psi}^\dagger(k)D(k)\boldsymbol{\psi}(k).
\end{equation}
The eigenenergies are given by $D(k)={\rm diag}(vk-t_\perp,-vk-t_\perp ,vk+t_\perp,-vk+t_\perp)$
where $v=\sqrt{v_F^2+\alpha_{SO}^2}$ is the renormalized Fermi velocity. These dispersion relations
are shown in Fig. \ref{fig:SPE}.

It is worth emphasising that while in the original basis $c_{\sigma,a}$, $\sigma$ corresponded to the physical spin and $a$ to the helical edge in question; in the new basis however $\psi_{\sigma,a}$, $\sigma$ corresponds to helicity and $a$ to the band index.  Thus the transformation matrix \eqref{matrix_rotation} encodes the relationship between spin and helicity that will be crucially important when potential disorder is added in section \ref{sec:disorder}.

\subsection{Interacting Hamiltonian in rotated basis}


In the new basis $\boldsymbol{\psi}$, the interaction part of the Hamiltonian becomes 

\begin{eqnarray}\nonumber\label{Ham_int_vector2}
 &&H_{\rm int}=\sum_{x,\sigma\sigma' aa'}\left(U_+{\psi}^\dagger_{\sigma a}(x){\psi}_{\sigma a}(x){\psi}^\dagger_{\sigma' a'}(x){\psi}_{\sigma' a'}(x)\right.\\\nonumber
 &&+\left.U_-\sum_{a_1,a_2}{\psi}^\dagger_{\sigma a}(x)(\tau^x)_{aa_1}{\psi}_{\sigma a_1}(x){\psi}^\dagger_{\sigma' a'}(x)(\tau^x)_{a'a_2}{\psi}_{\sigma' a_2}(x)\right).
\end{eqnarray}
with $U_\pm=(U_0\pm U)/2$.
In the continuum limit, it is convenient to rewrite the field operators in terms of slow modes near each Fermi point \cite{GiamarchiBook2003,FradkinBook2013,GogolinBook2004}:
\begin{eqnarray}\nonumber
 \psi_{+,1}(x)\rightarrow R_1(x)e^{ik_F^1x},\quad \psi_{-,1}(x)\rightarrow L_1(x)e^{-ik_F^1x}\\\nonumber
 \psi_{+,2}(x)\rightarrow R_2(x)e^{ik_F^2x},\quad \psi_{-,2}(x)\rightarrow L_2(x)e^{-ik_F^2x}.
\end{eqnarray}
Here the Fermi momenta $k^{1,2}_F$ are given by $(\epsilon_F\pm t_{\perp})/v$. The non-interacting Hamiltonian can be written in terms of the slow modes in a standard way
\begin{equation}\label{non_int}
 H_0=-iv\int dx\sum_{a}\left(R_a^\dagger\partial_xR_a-L_a^\dagger\partial_xL_a\right).
\end{equation}
The interaction Hamiltonian acquires the form
\begin{eqnarray}\label{fw1}
 H_{\rm int}&=&\frac{U_+}{2}\int dx\sum_a \left(R^\dagger_aR_a+L^\dagger_aL_a\right)^2\\\label{fw2}
 &+&U_-\int dx \left(R^\dagger_1R_1R^\dagger_2R_2+L^\dagger_1L_1L^\dagger_2L_2\right)\\\label{bw1}
  &+&U_-\int dx \left(R^\dagger_1R_2L^\dagger_1L_2+L^\dagger_2L_1R^\dagger_2R_1\right)\\\label{bw2}
  &+&U_-\int dx \left(R^\dagger_1R_2L^\dagger_2L_1e^{2i\Delta k_F\,x}+h.c.\right),
\end{eqnarray}
with $U_\pm=U_0\pm U$ and $\Delta k_F=k_F^1-k_F^2 = 2t\perp/v$.

\section{Bosonization and RG analysis}

To account for the effects of  the interaction,  it is natural to pass to the bosonic description of fermionic fields.
The fermionic fields are represented by the vertex operators
\begin{equation}
 R_i=\frac{\kappa_i}{\sqrt{2\pi a_0}}e^{i\sqrt{4\pi}\phi_i^R},\quad L_i=\frac{\kappa_i}{\sqrt{2\pi a_0}}e^{-i\sqrt{4\pi}\phi_i^L}\,.
\end{equation}
The bosonic fields satisfy the equal time commutation relations
\begin{eqnarray}
[\phi_i^R(x),\phi_j^L(x)]&=&\frac{i}{4}\delta_{ij}\quad(\mbox{same point}),\\
 \left[\phi_i^{\eta}(x),\phi_j^{\eta'}(y)\right]&=&\frac{i}{4}\eta\delta_{ij}\delta_{\eta\eta'}{\rm sgn}(x-y).
\end{eqnarray}
After bosonization \cite{GiamarchiBook2003,GogolinBook2004,FradkinBook2013}, the noninteracting part of the Hamiltonian (\ref{non_int}) combines with (\ref{fw1}) and 
(\ref{fw2}) into the quadratic bosonic Hamiltonian
\begin{eqnarray}
H_{\rm quad}&=& v\int dx\sum_a\left((\partial_x\phi_a^R)^2+(\partial_x\phi_a^L)^2\right)\\
&+&g\int dx\sum_a\left(\partial_x\phi_a^R+\partial_x\phi_a^L\right)^2\\
&+&g'\int dx\left(\partial_x\phi_1^R\partial_x\phi_2^R+\partial_x\phi_1^L\partial_x\phi_2^L\right).
\end{eqnarray}
Here $g=(U_0+U)a_0/2\pi$, $g'=(U_0-U)a_0/2\pi$ and $a_0$ the lattice constant. The interaction Hamiltonian also generates the backscattering terms
\begin{eqnarray}&&
 H_{\rm bs}=-\frac{g'}{\pi a_0^2}\int dx\left(e^{i\sqrt{4\pi}(\phi_1^L-\phi_1^R-\phi_2^L+\phi_2^R)}+h.c\right)\quad\quad\\&&
\,\,\,\,\,+\frac{g'}{\pi a_0^2}\int dx\left(e^{i\sqrt{4\pi}(\phi_1^L+\phi_1^R-\phi_2^L-\phi_2^R)}e^{i\Delta x}+h.c\right).
\end{eqnarray}
In terms of  the new bosonic fields
\begin{eqnarray}
 \begin{bmatrix}
      \varphi_{+}  \\
      \theta_{+} \\
      \varphi_{-}  \\
      \theta_{-} \\
 \end{bmatrix}=\frac{1}{\sqrt{2}}\begin{bmatrix}
     1 & 1 & 1 & 1  \\
     1 & -1 & 1 & -1  \\
     1 & 1 & -1 & -1  \\
     1 & -1 & -1 & 1  \\
 \end{bmatrix} \begin{bmatrix}
     \phi_1^L   \\
     \phi_1^R   \\
     \phi_2^L   \\
     \phi_2^R   \\
 \end{bmatrix},
\end{eqnarray}
the full Hamiltonian (\ref{non_int}-\ref{bw2}) splits into two commuting parts $H=H_++H_-$. The Hamiltonian $H_+$ is given by 
\begin{equation}
 H_+=\frac{u_+}{2}\int dx\left[\frac{(\partial_x\varphi_+)^2}{K}+(\partial_x\theta_+)^2K\right],
\end{equation}
where $u_+=\sqrt{(v+g')(v+g'+4g)}$ and the Luttinger parameter is $K=\sqrt{\frac{v+g'}{v+g'+4g}}$. The second part of  the 
Hamiltonian, $H_-$ is
\begin{eqnarray}\label{Ham_spin}
 H_-&=&\frac{u_-}{2}\int dx\left[(\partial_x\varphi_-)^2+(\partial_x\theta_-)^2\right]\\\nonumber
 &-&\frac{g'}{\pi a_0^2}\int dx \left(\cos(\sqrt{8\pi}\theta_-)-\cos(\sqrt{8\pi}\varphi_-+2\Delta k_F x)\right),
\end{eqnarray}
with $u_-=v-g'$. Note that due to the helical nature of the fermionic modes, the bare Luttinger 
parameter $K_-$ of the Hamiltonian $H_-$ equals unity. 
The RG equations depend on the ratio between the running scale  to the tunneling amplitude.

For energies above $t_\perp$, one can ignore the oscillating part $2\Delta k_F x$ in the second cosine, and the model is equivalent to the bosonized form of the XYZ chain, naturally tuned to be on a $Z_4$ plane in the phase diagram (see e.g. Ref.~\cite{GogolinBook2004}.   The RG equations are
\begin{eqnarray}
\label{eq:Z4}
 \frac{\partial K_-}{\partial \ell}=0,\quad \frac{\partial \tilde{g}}{\partial \ell}=0.
\end{eqnarray}
Here $\ell=\ln\Lambda_0/\Lambda$ (with $\Lambda$ being a running energy scale), 
and $\tilde{g}=g'/u_-=a_0(U_0-U)/2\pi u_-$. Clearly, neither the Luttinger 
parameter nor the amplitude of the cosines renormalize in this regime.  The Luttinger parameter therefore remains unity, and the theory remain gapless as be shown by refermionization back to the original fermionic degrees of freedom.

However, below the energy scale $t_\perp$,  the presence of the oscillations $2\Delta k_F x$ in the second cosine term in (\ref{Ham_spin} become important, and therefore averaged over long energy scales, this entire cosine term can be neglected in the RG flow at these energy scales \cite{Nersesyan1993}.  One is then left with the well known sine-Gordon model;
the RG equations in this case read \cite{GogolinBook2004,GiamarchiBook2003}
\begin{eqnarray}
 \frac{\partial K_-}{\partial \ell}=-\tilde{g}^2,\quad \frac{\partial \tilde{g}}{\partial \ell}=(1-K_-)\tilde{g}.
\end{eqnarray}
We see that both $K_-$ and $\tilde{g}$ always flow to strong coupling as the energy scale is reduced ($\ell\rightarrow \infty$).
Therefore the term $\cos(\sqrt{8\pi}\theta_-)$ opens a gap in the mode described by $\phi_-, \theta_-$.  
In this situation, the system flows to one of two strong coupling fixed points  depending on the sign of $g'$. We note however that the other mode, $\phi_+,\theta_+$ always remains gapless in the absence of any Umklapp scattering.

At this point, it is also worth emphasizing the importance of interchain hopping $t_\perp \ne 0$ in the above result.  In its absence $t_\perp=0$, one would never enter the second range of RG flow, and Eq.~\eqref{eq:Z4} would be valid until arbitrarily low temperatures.  Thus the interchain hopping is necessary for a strong coupling phase to occur (for weak interactions).
We now proceed to characterize the two strong coupling phases by looking at potential local order parameters.  As one of the modes remains gapless, these local order parameters are never non-zero in the thermodynamic limit, but rather the phase is identified as the order parameter with the slowest decaying correlations; see e.g. Refs.~\cite{Carr2013,Starykh2000}.

\subsection{Intra-mode interaction stronger than inter-mode interaction ($g'>0$)}

For positive $g'$ the minimum of $-g'\cos(\sqrt{8\pi}\theta_-)$ takes place at 
 $\theta_-=\sqrt{\frac{\pi}{2}}n$ with $n\in\mathbb{Z}$.
 In this case, the order parameter (with $\bar{k}_F= (k_F^1+k_F^2)/2 \equiv \epsilon_F/v$)
 \begin{eqnarray}
  \mathcal{O}_{I}&=&i(e^{2i\bar{k}_F x}\psi_{+,1}^\dagger \psi_{-,2}+e^{-2i\bar{k}_F x}\psi_{-,1}^\dagger \psi_{+,2} -h.c.)\\\nonumber
     &=&\frac{1}{\pi a_0}\left[\cos\sqrt{4\pi}(\phi_1^R+\phi_2^L)+\cos\sqrt{4\pi}(\phi_2^R+\phi_1^L)\right]\\
     &=&\frac{2}{\pi a_0}\cos(\sqrt{2\pi}\theta_-)\cos(\sqrt{2\pi}\varphi_+),
    \end{eqnarray}
 becomes dominant as $\langle\cos\sqrt{2\pi}\theta_-\rangle\neq0$. In terms of the original helical fermions $c_{\sigma,a}$,
 this order parameter reads
 \begin{eqnarray}\nonumber
  \mathcal{O}_{I}
  &=&\cos 2\bar{k}_F x\sum_{\sigma\sigma' aa'}c^\dagger_{\sigma a}(\tau^y)_{aa'}[\cos2\beta\sigma^x-\sin2\beta\sigma^z]_{\sigma\sigma'}c_{\sigma'a'}\\
  &-&\sin 2\bar{k}_F x\sum_{\sigma\sigma' aa'}c^\dagger_{\sigma a}(\tau^y)_{aa'}(\sigma^y)_{\sigma\sigma'}c_{\sigma'a'}.
 \end{eqnarray}
 To understand the structure of this order parameter, we can rotate the spin quantization axis in the $xz$ plane, 
 by defining the rotated fermionic operators $\tilde{c}_{\sigma,a}=\sum_{\sigma'}(e^{i(\beta-\pi/4)\sigma^y})_{\sigma\sigma'}c_{\sigma',a}$. In this basis
 \begin{equation}
 \label{eq:OPSN}
  \mathcal{O}_{I}=\sum_{\sigma\sigma' aa'}\tilde{c}^\dagger_{\sigma a}(\tau^y)_{aa'}[\cos 2\bar{k}_F x\;\sigma^z-\sin 2\bar{k}_F x\;\sigma^y]_{\sigma\sigma'}\tilde{c}_{\sigma' a'}
 \end{equation}
The $\tau_y$ in this order parameter means that a pattern of currents is flowing between the two spin edges.  The presence of $\sigma^x$ and $\sigma^y$ (rather than $\sigma^0$) means that these are spin currents; the spatially dependent part in brackets describes a spiral for the axis of quantization of these currents.  This order parameter therefore may be interpreted as a spin-nematic phase \cite{Nersesyan1991,GogolinBook2004,FradkinBook2013} 
in the spirally varying tilted spin basis.

\subsection{Inter-mode interaction stronger than intra-mode interaction ($g'<0$)}

 For  negative $g'$ the minimum of $-g'\cos(\sqrt{8\pi}\theta_-)$ occurs at $\theta_-=\sqrt{\frac{\pi}{2}}\left(n+\frac{1}{2}\right)$ with $n\in\mathbb{Z}$.
 In this case, the order parameter 
 \begin{eqnarray}
     \mathcal{O}_{II}&=&e^{2i\bar{k}_F x}(\psi_{+,1}^\dagger \psi_{-,2}+\psi_{+,2}^\dagger \psi_{-,1}) +h.c.\\\nonumber
     &=&\frac{1}{\pi a_0}\left[\sin\sqrt{4\pi}(\phi_1^R+\phi_2^L)-\sin\sqrt{4\pi}(\phi_2^R+\phi_1^L)\right]\\
     &=&\frac{2}{\pi a_0}\sin(\sqrt{2\pi}\theta_-)\cos(\sqrt{2\pi}\varphi_+),
    \end{eqnarray}
 becomes dominant as $\langle\sin\sqrt{2\pi}\theta_-\rangle\neq0$. In terms of the original helical fermions $c_{\sigma,a}$,
 this order parameter reads
 \begin{eqnarray}\nonumber
  \mathcal{O}_{II}
  &=&\cos 2\bar{k}_F x \sum_{\sigma\sigma' aa'}c^\dagger_{\sigma a}(\tau^z)_{aa'}[\cos2\beta\sigma^x-\sin2\beta\sigma^z]_{\sigma\sigma'}c_{\sigma'a'}\\
  &-&\sin 2\bar{k}_F x \sum_{\sigma\sigma' aa'}c^\dagger_{\sigma a}(\tau^z)_{aa'}(\sigma^y)_{\sigma\sigma'}c_{\sigma',a'}.
 \end{eqnarray}
 Again, performing a rotation in the spin basis $\tilde{c}_{\sigma,a}=\sum_{\sigma'}(e^{i(\beta-\pi/4)\sigma^y})_{\sigma\sigma'}c_{\sigma',a}$ the order
 parameter can be written in the familiar form
 \begin{equation}
  \mathcal{O}_{II}=\sum_{\sigma\sigma' aa'}\tilde{c}^\dagger_{\sigma a}(\tau^z)_{aa'}[\cos 2\bar{k}_F x\,\sigma_z-\sin 2\bar{k}_F x\,\sigma^y]_{\sigma\sigma'}\tilde{c}_{\sigma',a'}
 \end{equation}
The only difference between this order parameter and that in Eq.~\eqref{eq:OPSN} is the replacement of $\tau^y$ with $\tau^x$.  This means that instead of spin currents, one has a pattern of spins, with the two different helical edges antiferromagnetically connected.   We can therefore interpret this order parameter as a  spin density wave, where as before the axis of quantization traces a spiral pattern along the edge of the sample.

\begin{figure}
	\centering
		\includegraphics[scale=0.8]{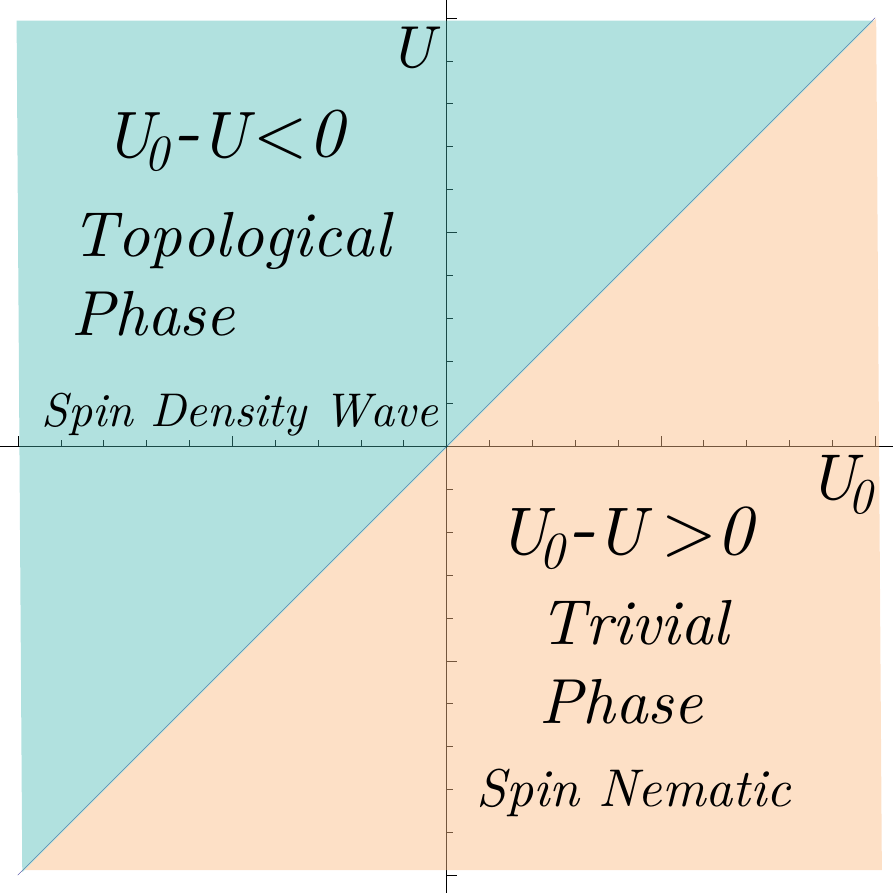}
	\caption{(color online).- Dominant order parameter for two interacting helical modes, for
	strong tunneling. Here $g'=(U_0-U)/2\pi$, where $U_0$ is the interaction strength within an helical mode, whereas $U$ is the mutual interaction
	between the modes. Above the diagonal the dominant correlations are of the spin-density wave type; while below the diagonal, the dominant correlations are of the spin-nematic type.   $H_-$ remain gapless along the diagonal $U_0=U$.}\label{fig:PD}
\end{figure}

Putting these two results together, the entire phase diagram of the problem in the absence of the disorder 
is depicted in Fig.~\ref{fig:PD}.

\section{Disorder}
\label{sec:disorder}

We now consider the response of the coupled edge system to backscattering; first we consider the case of single impurities and then we go on to look at random disorder.  We follow closely methods previously developed for ordinary (non-helical) two-leg ladders \cite{Starykh2000,Carr2011,Carr2013}.  Rather similarly to two-leg ladders, we will find one of the strong coupling phases is particularly susceptible to localization by disorder; while in the other phase, the system remains a ballistic conductor, even when disorder is added (rather like the original helical edges before they were coupled).

We then go on to show that the conducting phase actually has emergent topological properties, namely zero-energy boundary states, before discussing experimental signatures of the results of the calculations in this section.

\subsection{Single impurity}
For isolated helical modes, non magnetic impurities cannot localize the metallic state for moderate
interaction. In the non-interacting limit   the backscattering between
counter-propagating modes is not allowed by the Kramers theorem. 
If interaction within  a helical mode is strong ($K<1/4$) the single impurity is a relevant perturbation.
For the random disorder the   localization occurs at $K<3/8$) \cite{Schmidt2012,Kainaris2014}. 
Here we analyze the fate of the conducting state when two helical modes are present. 
The presence of a non magnetic impurity at $x=0$ generates the scattering  processes (here $c_{\sigma,a}=c_{\sigma,a}(x=0)$)	
\begin{eqnarray}
 H_{\rm imp}^\parallel&=&\sum_{\sigma,a}\mu^\parallel_{a}c_{\sigma,a}^\dagger c_{\sigma,a},\\
 H_{\rm imp}^\perp&=&\mu^\perp(c_{\uparrow,1}^\dagger c_{\uparrow,2}+c_{\downarrow,2}^\dagger c_{\downarrow,1})+ h.c.
\end{eqnarray}
In general, for a TR invariant impurity potential, $\mu_{1,2}^\parallel$ are real numbers while $\mu^\perp=\mu_\Re^\perp+i\mu_\Im^\perp$
can be a complex number. A finite imaginary part of $\mu^\perp$ implies a breaking of an inversion symmetry by  disorder potential. Let 
us note that inversion symmetry is broken already at the  level of  the  single particle   Hamiltonian (\ref{H_0_nonint}), as the 
helical modes break explicitly the right-left symmetry due to their different spin projections.
Writing  the real and imaginary parts of $\mu^\perp$, the impurity  scattering processes $H_{\rm imp}^\perp$ read
\begin{eqnarray}\nonumber
H_{\rm imp}^\perp&=&\mu_\Re^\perp\sum_{\sigma\sigma'aa'}c_{\sigma a}^\dagger (\tau^x)_{aa'}\delta_{\sigma\sigma'}c_{\sigma' a'}\\
&-&\mu_\Im^\perp\sum_{\sigma\sigma'aa'}c_{\sigma a}^\dagger (\tau^y)_{aa'}(\sigma^z)_{\sigma\sigma'}c_{\sigma' a'}.
\end{eqnarray}
In the basis (\ref{psi}) the forward part of the impurity scattering is given by
\begin{eqnarray}\label{Imp_forward}&&
H^{\rm f}_{\rm imp}= \sum_{\sigma,a}\mu_a\psi_{\sigma,a}^\dagger \psi_{\sigma,a}
+\mu^\parallel_-\sum_{\sigma}(\psi^\dagger_{\sigma,1}\psi_{\sigma,2}+h.c)\nonumber \\&&
+\mu_\Im^\perp\cos2\beta\sum_{\sigma\sigma'aa'}\psi_{\sigma a}^\dagger (\tau^y)_{aa'}(\sigma^z)_{\sigma\sigma'}\psi_{\sigma' a'}\,,
\end{eqnarray}
with $\mu_{1,2}=\frac{\mu^\parallel_1+\mu^\parallel_2}{2}\pm\mu_\Re^\perp$ and $\mu_-^\parallel=\frac{\mu^\parallel_1-\mu^\parallel_2}{2}$.	
The  backscattering term are accounted by
\begin{eqnarray}\nonumber
H^{b}_{\rm imp}=
\mu^\perp_{\Im}\sin2\beta\sum_{\sigma\sigma' aa'}\psi^\dagger_{\sigma a}(\tau^y)_{aa'}(\sigma^x)_{\sigma\sigma'}\psi_{\sigma' a'}.
\end{eqnarray}
The forward processes do not play any role in the Anderson localization and therefore will be neglected.
One is left with  the backscattering term
\begin{equation}
H_{\rm imp}^{\rm b}=\mu^\perp_\Im\sin2\beta  \sum_{\sigma\sigma' aa'}\psi^\dagger_{\sigma a}(\tau^y)_{aa'}(\sigma^x)_{\sigma\sigma'}\psi_{\sigma' a'}.
\end{equation}
After  bosonization, this term reads
\begin{eqnarray}\nonumber
 H_{\rm imp}^{\rm b}&=&-i\mu^\perp_\Im \sin2\beta\left[R_1^\dagger L_2+L_1^\dagger R_2-R_2^\dagger L_1-L_2^\dagger R_1 \right]\\
 &=&-\frac{2\mu^\perp_\Im}{\pi a_0} \sin2\beta\cos(\sqrt{2\pi}\theta_-)\cos(\sqrt{2\pi}\varphi_+).
\end{eqnarray}
For $g'>0$ when the system is in the spin-nematic phase, the expectation value of $\cos\sqrt{2\pi}\theta_-$ is finite. This implies that the scattering operator is 
determined by $\cos\sqrt{2\pi}\varphi_+$. Under RG its scaling dimension is $K/2<1$.
Therefore it is a relevant perturbation, making the system an insulator.
In the opposite case, for $g'<0$ in the spin-density wave state, the expectation value of $\cos\sqrt{2\pi}\theta_-$ is zero and 
the backscattering operator is always irrelevant.
Therefore the system remains conducting.

\subsection{Random disorder}
We now turn to another limit of disorder, where one considers many weak non-magnetic impurities.  In this case, the previous analysis should be modified. 
The disorder is accounted by  
$H_{\rm dis}=H_{\rm dis}^\parallel+H_{\rm dis}^\perp$ with
\begin{eqnarray}
 H_{\rm dis}^\parallel&=&\int dx \sum_{\sigma,a}\mathcal{U}_a^\parallel(x)c_{\sigma,a}^\dagger(x) c_{\sigma,a}(x),\\\nonumber
 H_{\rm dis}^\perp&=&\int dx\mathcal{U}^\perp(x)(c_{\uparrow,1}^\dagger(x) c_{\uparrow,2}(x)+c_{\downarrow,2}^\dagger(x) c_{\downarrow,1}(x))+ h.c.
\end{eqnarray}
The components of random potential $\mathcal{U}^\perp(x')$ are given by
\begin{eqnarray}&&
 \mathcal{U}^\perp(x)=\int dy U(x,y)\chi_1(y)\chi^*_2(y-d),\\&&
\mathcal{U}^\parallel_a(x)=\int dy U(x,y)\chi_a(y)\chi^*_a(y),
\end{eqnarray}
where $U(x,y)$ is the two dimensional random potential generated by the impurities. The wavefunction of the the helical mode $a$ in the direction perpendicular to the motion is $\chi_a(y)$.
The function $\chi(y)$ is peaked around zero, as the helical edge modes are quasi-onedimensional. The separation $d$ between
the modes is assumed to be constant. 

We assume that the disorder  at different points is  uncorrelated, i.e.   $\overline{\mathcal{U}_a^\parallel(x)\mathcal{U}_b^\parallel(x')}=\delta_{ab}\delta(x-x')$
and $\overline{(\mathcal{U}^\perp(x))^*\mathcal{U}^\perp(x')}=D\delta(x-x')$. As in the case of single impurity, the disordered
Hamiltonian $H_{\rm dis}^\parallel$ in the single-particle diagonal basis $\psi$ contains just forward scattering terms, that
do not localize the system. We concentrate in $H_{\rm dis}^\perp$ which in the $\psi$ basis becomes
\begin{eqnarray}\nonumber
H^\perp_{\rm dis}&=&\int dx\sum_{\sigma\sigma' aa'}\left(\mathcal{U}_\Re^\perp(x)\psi_{\sigma a}^\dagger(x) (\tau^z)_{aa'}\delta_{\sigma\sigma'}\psi_{\sigma a'}(x)\right.\\
&+&\left.\mathcal{U}_\Im^\perp(x)\psi_{\sigma a}^\dagger(x) (\tau^y)_{aa'}(\hat{n}(\beta)\cdot\vec{\sigma})_{\sigma\sigma'}\psi_{\sigma' a'}(x)\right).
\end{eqnarray}
where $\mathcal{U}_{\Re(\Im)}^\perp$  is the real (imaginary) part of the disorder potential $\mathcal{U}^\perp$. The vector 
$\vec{n}$ is unitary and explicitly given by $\vec{n}(\beta)=(\sin2\beta,0,\cos2\beta)$. Focusing on the backscattering terms we have
\begin{equation}
H_{\rm dis}^{\rm b}=\frac{1}{2}\int dx\sum_{\sigma\sigma' aa'}\eta(x,\beta)\psi^\dagger_{\sigma a}(\tau^y)_{aa'}(\sigma^x)_{\sigma\sigma'}\psi_{\sigma' a'},
\end{equation}
with $\eta(x,\beta)=2\mathcal{U}_\Im^\perp(x)\sin2\beta$. Under bosonization, this term becomes
\begin{eqnarray}\label{disorder_backscattering}
 H_{\rm imp}^{\rm b}&=&-\frac{i}{2}\int dx\eta(x,\beta)\left[(R_1^\dagger L_2-R_2^\dagger L_1)e^{-i\delta x}-h.c \right]\\\nonumber	
 &=&-\frac{1}{\pi a_0}\int dx\eta(x,\beta)\cos(\sqrt{2\pi}\theta_-)\cos(\sqrt{2\pi}\varphi_++\delta x),
\end{eqnarray}
with $\delta=2\epsilon_F/v$. Averaging over disorder, one finds the replicated action that is generated  by  the backscattering term 
(\ref{disorder_backscattering}) 
\begin{eqnarray}\nonumber
 S_{\rm imp}^{b,AV}&=&\frac{D}{(\pi a_0)^2}\sum_{\alpha\beta}\int dx d\tau_1d\tau_2\cos(\sqrt{2\pi}\theta_{-}^\alpha)\cos(\sqrt{2\pi}\varphi_+^\alpha)\\
 &\times&\cos(\sqrt{2\pi}\theta_{-}^\beta)\cos(\sqrt{2\pi}\varphi_+^\beta).
\end{eqnarray}

Deep in the gapped phase we can expand $\cos(\sqrt{2\pi}\theta_{-})$ around its minimum $\theta_{-}=\theta_{\rm min}+\delta\theta$. 
Integrating the massive $\delta\theta$ mode, the model for the charge field $\varphi_+$ maps to a Giamarchi-Schultz 
\cite{Giamarchi1988} model with Luttinger parameter $K'=2K$. Therefore the random disorder is a relevant
perturbation for $K<3/4$. 

\subsection{Conductance as a function of temperature}

\begin{figure}
	\centering
		\includegraphics[scale=0.55]{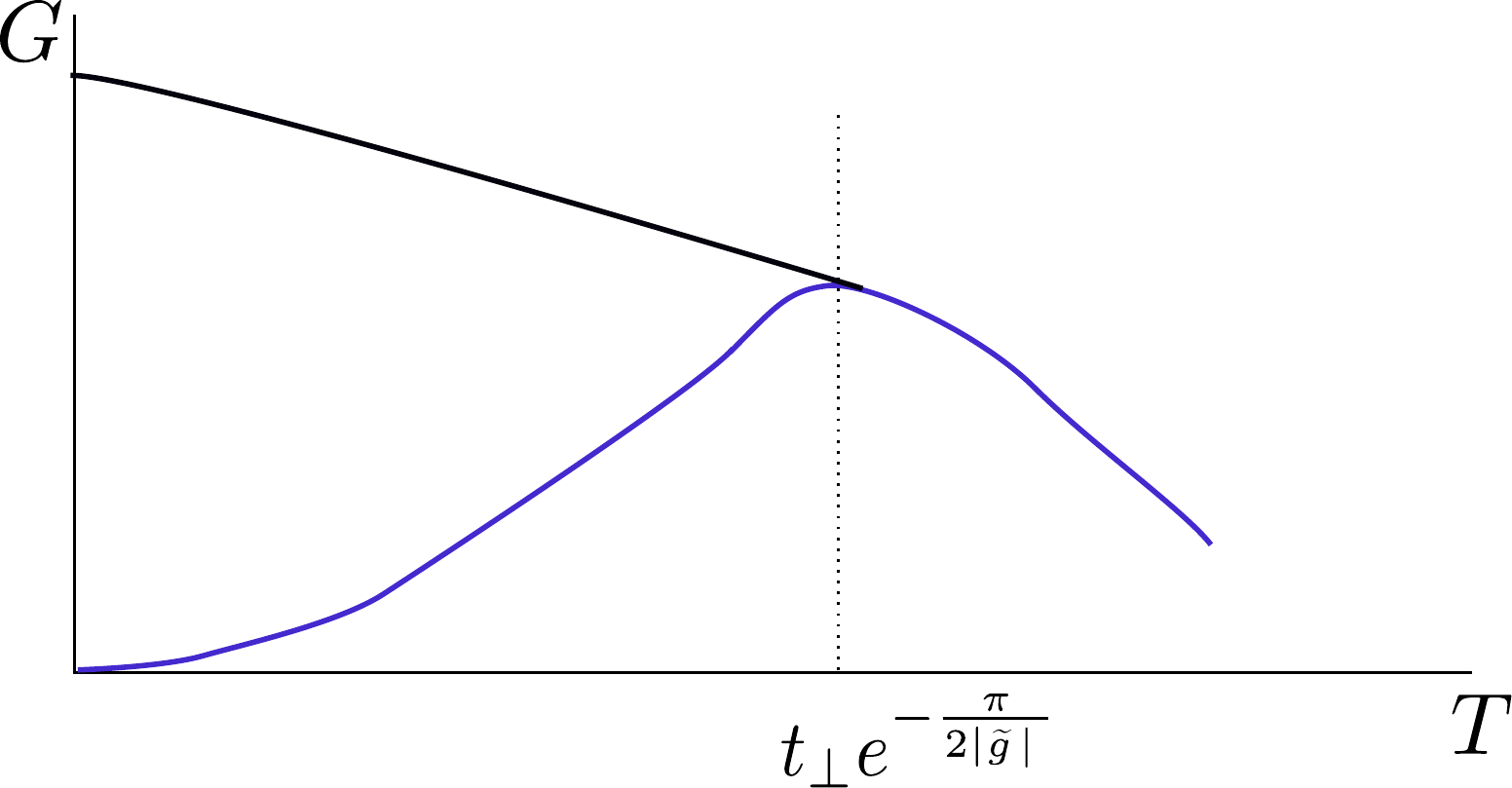}
	\caption{(color online) Non-monotonic conductance as a function of temperature for moderate interactions. At energies below the gap $\Delta_{\rm edge} \sim t_\perp \exp({-\frac{\pi^2u_-}{a_0|U-U_0|}})$
the conductance renormalizes to zero in the presence of small concentration of impurities (blue curve)
if $\tilde{g}=a_0(U_0-U)/2\pi u_->0$ (spin-nematic phase) .  For $\tilde{g}=a_0(U_0-U)/2\pi u_-<0$ (spin-density wave phase), the conductance remains finite (black curve). At energies above the gap, the conductance is
dominated by single particle tunneling.
	}\label{fig:conductance}
\end{figure}

At energy scales above $\Delta_{\rm edge} \sim t_\perp e^{-\pi/2|\tilde{g}|}$, the conductance is dominated by the single particle tunneling \cite{Santos2015}. For the moderate interaction strength this perturbation is  irrelevant,
and conductance increases with lowering the temperature.  
Below the energy scale $\Delta_{\rm edge}$, two-particle processes are dominant, and open the gap in the spin sector.
In the topological phase ($g'<0$) this gap protects the conducting mode (for the charge sector)  
in the presence  of a single impurity,  while  in topologically trivial phase ($g'>0$) it does not. 
For the random disorder the topological phase remains conducting for $K>3/4$.
A schematic  dependence of  conductance on temperature is shown in Fig. \ref{fig:conductance}. 

\subsection{Boundary zero modes in the protected phase}

\begin{figure}
	\centering
		\includegraphics[scale=0.55]{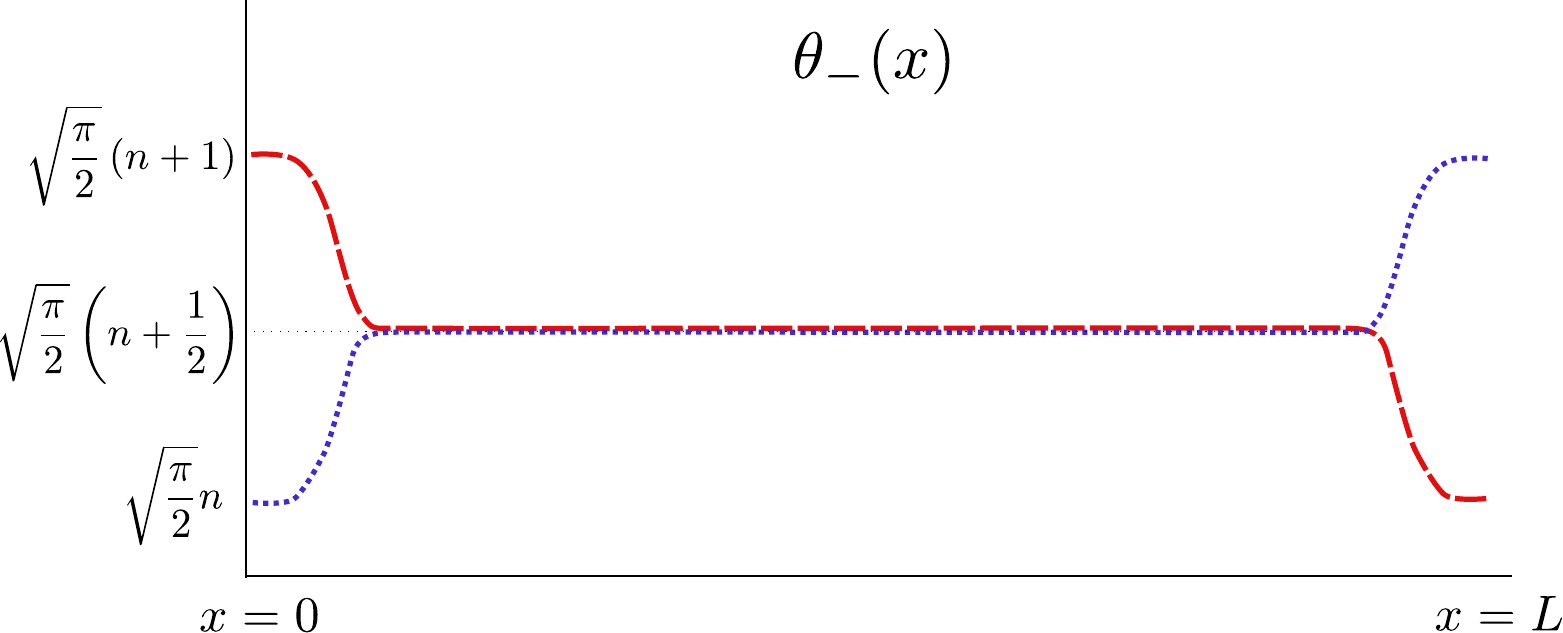}
	\caption{(color online). Spatial profile of the $\theta_-(x)$ field in the topological phase $(g'<0)$.
	The two different groundstates in a finite helical system correspond to the
	two choices of kink anti-kink in the boundary, where the field  has to minimize
	the backscattering potential. Different colors represent different ground state profiles for $\theta_-(x)$.
	}\label{fig:Theta}
\end{figure}

In order to reveal the existence of zero modes, we introduce a strong non magnetic impurity that pinches off a section of the helical modes. This discussion is then analogous to the one presented in \cite{Keselman2015,Kainaris2015} for the case of non-helical chains.
These impurities are modeled by 
\begin{eqnarray}\label{potential_well}\nonumber
 U_{\rm well}&=&ih_w\sum_{a,i}(R_a^\dagger L_{\bar{a}}+L_{\bar{a}}^\dagger R_a)\delta(x-x_i)+h.c.,\\
&=&\left.\frac{2h_w}{\pi a_0}\cos(\sqrt{2\pi}\varphi_+)\cos(\sqrt{2\pi}\theta_-)\right|_{x=0}^{x=L},
\end{eqnarray}
where $x_1=0$ and $x_2=L$ and $\delta$ a Dirac delta function. The backscattering strength $h_w$ is assumed to be larger than
any other relevant energy scale in the problem. The potential well (\ref{potential_well}) pins the field $\theta_-$ to the value $\sqrt{\frac{\pi}{2}}m $ with $m\in\mathbb{Z}$, close to the boundary. In the bulk the field $\theta_-$ is pinned to either 
$\sqrt{\frac{\pi}{2}}n$ for $g'>0$ or 
$\sqrt{\frac{\pi}{2}}(n+1/2)$ for $g'<0$. This implies that for $g'<0$ the field $\theta_-$ has to 
change by $\pm\frac{1}{2}\sqrt{\frac{\pi}{2}}$ close to the boundary (see Fig. \ref{fig:Theta}). This kink in the $\theta_-(x)$ field corresponds to a spin $1/4$ excitation  near  the edge. 
The two different ground states correspond to configurations with  kink and anti-kink pairs 
that are shown  in Fig. \ref{fig:Theta}.
Both configurations have the same energy.   This degeneracy of the $\theta_-$ field at the edge of the samples allows particles to tunnel in or out at the edges without paying the energy cost of the gap.  One may therefore describe these modes as topologically protected localised zero-mode at the boundaries of the sample \cite{Keselman2015,Kainaris2015}.

As we discussed in the previous section, for $g'<0$ the system is protected against localization by single impurity due to the existence of the spin gap. Therefore the spin density wave phase in this model is indeed topological, being protected against single impurity backscattering and hosting
fractionalized zero modes on its ends. By the same analysis, we find that the spin nematic phase is topologically trivial.

\subsection{Experimental signatures}
There are several predictions we have made  that can be tested experimentally.
First, the electric conductance studied  above may  be measured in two terminal experiment.
In this measurement one  attaches  ohmic leads on the edge of the sample,  as shown in Fig \ref{fig:Exp_leads}. Our theory predicts the dependence of the two terminal conductance on temperature,
see Fig.\ref{fig:conductance}.

\begin{figure}
	\centering
		\includegraphics[width=\linewidth]{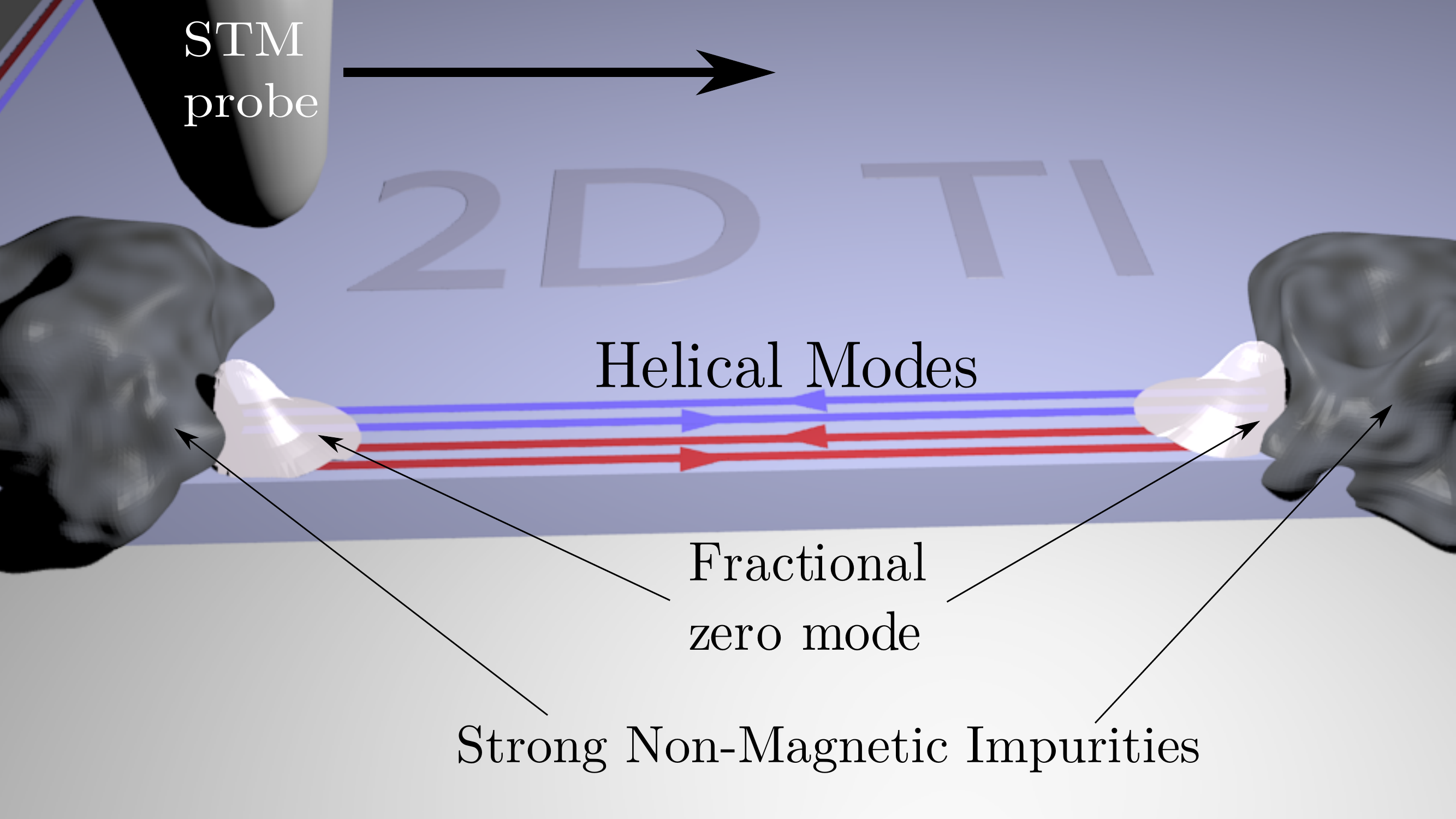}
	\caption{(color online) Experimental setup for the detection of fractionalized zero modes. 
	By moving the  tip of STM  parallel  to the edge  one measures the tunneling current as a function of a distance from the end point.(See text).
	}\label{fig:Exp}
\end{figure}

Another type of experimental  study involves Scanning Tunneling microscope (STM). 
We propose to perform such experiment after adding two  non-magnetic impurities to the system.
Provided that the amplitude of impurities are bigger that the size of the gap in bulk ($\Delta_{\rm bulk}$), a finite  part of the helical mode is cut off the  rest of the system. 
If the system is in topologically  non trivial phase we expect to find  fractional zero-energy 
modes at the end points of the constriction, see Fig. \ref{fig:Exp}.
By scanning the tip of the tunneling microscope away from the end points one expected to see
a hard gap in the density of states of the size $\Delta_{\rm edge}$.
The tunneling density of states in the topological phase scales as \cite{Starykh2000}

\begin{equation}
 \nu(\epsilon)/\nu_0\propto	 \left\{
\begin{array}{ll}
      \left(\frac{\epsilon}{\Lambda}\right)^{\frac{1}{2K}-1}, &  \mbox{close to the strong impurities} \\
\\
\\
      \theta(\epsilon-\Delta_{\rm edge}), &  \mbox{away from the impurities} 
\end{array} 
\right.
\end{equation}
where $\nu_0$ is  a bare value of the density of states,  $\epsilon$ is the energy of the tunneling electron with respect to the Fermi energy, and $\Lambda$ is the ultra-violet cutoff.  
At  low  bias, the tunneling current has a power law zero bias anomaly  near the end points  (see Fig \ref{fig:Exp}),  where the  spin gap vanishes. Along the edge, but away from the impurities, the tunneling density of states $\nu(\epsilon)$ vanishes at bias smaller than  $\Delta_{\rm edge}$.

\section{Conclusions}

\begin{table}
\begin{tabular}{| l | c | c |}
  \hline			
  & $U_0>U$ & $U_0<U$ \\ 
  \hline
  Topological Protection  & No & Yes \\
\hline
& &\\
  Order Parameter & Spin Nematic & Spin Density Wave \\
& & \\
\hline
  Zero modes & No & Yes \\
  \hline  
\end{tabular}
\caption{Phases of two interacting helical modes. The  interaction strength within the same helical state is $U_0$ and  between different helical modes is $U$.}
\label{table:summary}
\end{table}

In this paper we studied the low energy physics of two helical edge modes,  
coupled by  tunneling and  electron-electron interaction.  
Our results are summarized in Table \ref{table:summary}.

We showed that the tunneling between the modes, in the presence of repulsive interaction 
and generic spin-orbit interaction, leads to the development of a spin gap. 
If  the interaction between Kramers partners  is stronger 
than the interaction between  states not connected by TR symmetry the system is topologically trivial.
The inclusion of weak  non-magnetic impurities localizes the conducting mode.
The two terminal conductance is a non monotonous function of temperature.

In the opposite limit,  the system is in topologically non-trivial phase.
The gap in the spin sector  protects the conducting phase against  backscattering 
by weak non-magnetic impurities. 
The protected phase has a ground state degeneracy and possess fractionalized zero energy edge-modes.
The later can be observed in tunneling spectroscopy experiments.
The two terminal conductance monotonously grows with decreasing  the temperature, reaching
$2e^2/h$ value at zero temperature.

\medskip

The authors acknowledge discussion with Y. Gefen,  N. Kainaris, A.D. Mirlin  and E. Sela.
This work has been supported by Israel Science Foundation (grant 584/14),
German Israeli Foundation  (grant 1167-165.14/2011) and the Israeli Ministry of Science.

\end{document}